\documentclass[numerical]{revtex4-1}
\usepackage{graphicx}
\usepackage{bm}
\draft 

\begin{document}

\def\Journal#1#2#3#4{{#1 }{\bf #2, }{ #3 }{ (#4).}} 
 
\def\BiJ{ Biophys. J.}                 
\def\Bios{ Biosensors and Bioelectronics} 
\def\LNC{ Lett. Nuovo Cimento} 
\def\JCP{ J. Chem. Phys.} 
\def\JAP{ J. Appl. Phys.} 
\def\JMB{ J. Mol. Biol.} 
\def\JPC{ J. Phys: Condens. matter}
\def\CMP{ Comm. Math. Phys.} 
\def\LMP{ Lett. Math. Phys.} 
\def\NLE{{ Nature Lett.}} 
\def\NPB{{ Nucl. Phys.} B} 
\def\PLA{{ Phys. Lett.}  A} 
\def\PLB{{ Phys. Lett.}  B} 
\def\PRL{ Phys. Rev. Lett.} 
\def\PRA{{ Phys. Rev.} A} 
\def\PRE{{ Phys. Rev.} E} 
\def\PRB{{ Phys. Rev.} B} 
\def\PD{{ Physica} D} 
\def\ZPC{{ Z. Phys.} C} 
\def\RMP{ Rev. Mod. Phys.} 
\def\EPJD{{ Eur. Phys. J.} D} 
\def\SAB{ Sens. Act. B} 
\title{Opsin vs opsin: new materials for biotechnological applications } 



\author{Eleonora Alfinito}
\email{eleonora.alfinito@unisalento.it}
\homepage{http://cmtg1.unile.it/eleonora1.html}
\affiliation{Dipartimento di Ingegneria dell'Innovazione,
Universit\`a del Salento, via Monteroni, I-73100 Lecce, Italy, EU}
\affiliation{CNISM,  Via della Vasca Navale, 84 - 00146 Roma, Italy. EU }
\author{Lino Reggiani}
\email{lino.reggiani@unisalento.it}
\affiliation{Dipartimento di Matematica e Fisica, \textit{Ennio de Giorgi},
Universit\`a del Salento, via Monteroni, I-73100 Lecce, Italy, EU}
\affiliation{CNISM,  Via della Vasca Navale, 84 - 00146 Roma, Italy, EU}

\date{\today}

\begin{abstract}
The need of new diagnostic methods satisfying, as an early detection, a low invasive procedure and a cost-efficient value,  is orienting the technological research toward the use of bio-integrated devices, in particular bio-sensors. 
The set of know-why necessary to achieve this goal is wide, from biochemistry to electronics and is summarized in an emerging  branch of electronics, called \textit{proteotronics}. 
Proteotronics is here here applied to state a comparative analysis of the electrical responses coming from type-1 and type-2 opsins. In particular, the procedure is used as an early investigation of a recently discovered family of opsins, the proteorhodopsins activated by blue light, BPRs. 
The results reveal some interesting and unexpected similarities between proteins of the two families, suggesting the global electrical response are not strictly linked to the class identity.
\end{abstract}

\pacs{87.14.et        
87.15.-v        
87.15.Pc        
87.19.R-                 
}

\maketitle 

\section{Introduction}
\label{Introduction}
Among the new frontiers of electronics we focus on the integration of biological matter
(specifically, proteins) into standard nanodevices. 
This is particularly useful for the development of devices to be used as non-invasive biological sensors, able to detect with high specificity and selectivity the presence of  drugs, toxins and also to act as cancer markers. 
To produce effective drugs and formulate novel strategies for medical therapies, a deep knowledge of the physical and chemical mechanisms correlating the protein structure  with its function is necessary.
In this perspective, the investigation of the physical properties of proteins, in particular the electrical properties,
is becoming more and more relevant. 
As matter of fact, the arising strategies to contrast many diffused illness like cancer \cite{Lee12} and brain diseases\cite{Lima05,*Deisseroth10} point toward a selective, focused and low-invasive action (the so-called targeted therapy).
This is made possible through the use of a precise procedure, acting at the molecular level.
\par  
Very recently a breakthrough method for controlling tissue and brain activity
in freely animals has been introduced. 
It is called \textit{optogenetics} \cite{Lima05,*Deisseroth10} and works
by genetically modifying neurons with the inclusion of light sensitive proteins
by prokaryotes, particularly channel-rhodopsins \cite{Hegemann10}. 
Then, the irradiation with light of specific wave lengths is able to control the tissue activity.
Furthermore, since the seminal paper by  Humayun and coworkers \cite{Humayun96} the activity of proteins sensitive to light is monitored, with the aim to use them for curing blindness. \cite{Ran13, *Saeedi11}
This investigation has followed the way of using  both organic materials \cite{Ghezzi11, *Ghezzi13} and living opsins 
\cite{Ahuja11}, and finally, the most recent result is the realization of a retinal prostheses. 
\par
Finally, this extremely versatile kind of proteins has found a primary role  in the sector of green and renewable energy  \cite{King12,Renu14}.
Conventional solar cells use  bulk inorganic materials like semiconductors, assembled to form a p-n junction. 
The efficiency is quite good \cite{Green12} but costs remain high, preventing their the large-scale diffusion. 
Recent studies point to overcome the cost limit by using nanostructured inorganic matter, mainly nanowires \cite{Garnett10, *Lovergine11}. 
At present, the results are interesting although the efficiency is quite low.  
Therefore, some researches point now toward the use of proteins, specifically bacteriorhodopsin, in solar cells of new generation.
These opsin-integrated devices which at present are only  proof of concept, promise many attactive features like the natural nanosized dimension, but also a fast response, higly efficience and potentially low cost \cite{King12,Renu14}. 
Among the recent proposals, the most innovative is a 
device consisting of a light absorbing surface composed of bR mutants on a thin layer of gold to produce ballistic electrons for photocurrent \cite{King12,Renu14}.
 
\par
The knowledge of sensing proteins like opsins is still quite incomplete,
both concerning those pertaining to the proton pump family (type-1 opsins), and those belonging to the GPCR family (type-2 opsins), like mammal proteins.
The operating principles are different for these proteins: type-1 opsins  pump protons outside the cellular membrane, while type-2 opsins activates an auxiliary protein called G-protein.
Furthermore, the 3D structure of type-1 and type-2
opsins, with seven transmembrane helices, is very similar.
This arises
the question of a possible interchange of these proteins in specific
applications, i.e. when the expected responses are more related to the topological
properties than to the functional properties. 
As matter of fact, it has to be emphasized that opsins by prokaryotes are more easy to be produced and used in vitro \cite{Jin06,Casuso07,Melikyan11}, with respect to proteins by eukaryotes \cite{Hou06, *Hou07}, and this of course is of interest for applications.  
\par
The investigation techniques concerning the opto-electrical properties of opsins are usually performed both in vivo and in vitro, ever revealing
interesting electrical properties.
Collecting all the researchs concerning
the application of electronic methods in biology is challenging topic  and it has originated a new branch of molecular electronics, the so called proteotronics \cite{proteotronics}.
One of the main objectives of {proteotronics} is to provide microscopic models for the physical properties of proteins: this step is preliminary to their exploitation in electronic devices. 
To achieve this scope, the complex biochemical mechanisms which rule the protein activity have to be translated into protocols of simple use for electronic applications. 
To get this result, it is possible to produce a physical map of the complex protein activity by using a network analogue. 
As a matter of fact, we cannot say where or when an assembly of atoms, like a protein, becomes an autonomously working object, a biomachine.
Nevertheless, the collective modes which underly the protein activity can be captured by using a quite standard approach in physics: it consists in the description of the \textit{interactions} inside the matter instead of the matter itself.  
%
%
There are very different ways to model interactions, here we use an analogous model  network approach: a set of nodes (matter) becomes a network, i.e. a system able to collectively operate, when the links among nodes (interactions)  are located, and a simple evolution law is assigned. 

The aim of this paper is to give a taste of how proteotronics works: getting
the crystallographic data, sketching some topological properties ,elaborating the expected electrical responses, suggesting novel experiments and applicative
uses.
 The tool used for these investigations, elsewhere called INPA, impedance network protein analogue, has been successfully tested to describe the AC and DC responses of some GPCRs and fine-tuned on bacteriorhodopsin.

The paper is organized as follows: Section II summarizes some relevant properties of opsins; Section III shortly illustrates the INPA model. Main results are collected in Section IV and Section V sketches the main conclusions.
\section{Opsins} 
Opsins are proteins able to convert light in energy useful for the
hosting cell. 
They are found in all the prokaryotes  (type-1) and mammals (type-2) and share a similar 3D structure (7 transmembrane $\alpha$-helices).
\par
In the group of  type-1 opsins the most studied is the bacteriorhodopsin, bR, and in the group of type-2 opsins  the counterpart is the bovine rhodopsin,  BR.
About 15 years ago, Beja and coworkers \cite{Beja01} identified a new type-1 opsin,
which was called proteorhodopsin, mainly sensitive to green light, GPR.
The relevance of this protein is enormous in ecology, since it is present
in all marine bacteria and its presence in different oceans is related to the water temperature.
Furthermore, the possible technological applications, the wide diffusion
of this protein and the ease of expression, for example in \textit{E. coli},
make it a kind of template for experimental and theoretical investigations.
Finally, in 2013 a new class of proteorhodopsins, sensitive to blue light,
BPR, was identified \cite{Ran13}. The discovery of these proteins open the door to novel technological applications, like optogenetics which specifically uses blue light.
Finally, the crystallographic data were published for three variants of this protein: 
a wild type, found in the Mediterranean sea, whose PDB entry is 4JQ6, and two mutants, found in the Pacific Ocean, whose PDB entries are 4KLY and 4KNF \cite{Berman00}. 
The most relevant difference among these proteins can be found in the quaternary structure which appears organized in hexamers in the wild type and in pentamers in the mutants. Furthermore, in the mutants, specific interconnections between all the couples of neighbouring protomers (single protein) are found.  The role of these interconnections is not completely cleared, but it seems  relevant for the functioning of the biological molecule. 
%
\begin{figure}
\centering
                \includegraphics[width=0.40\textwidth]{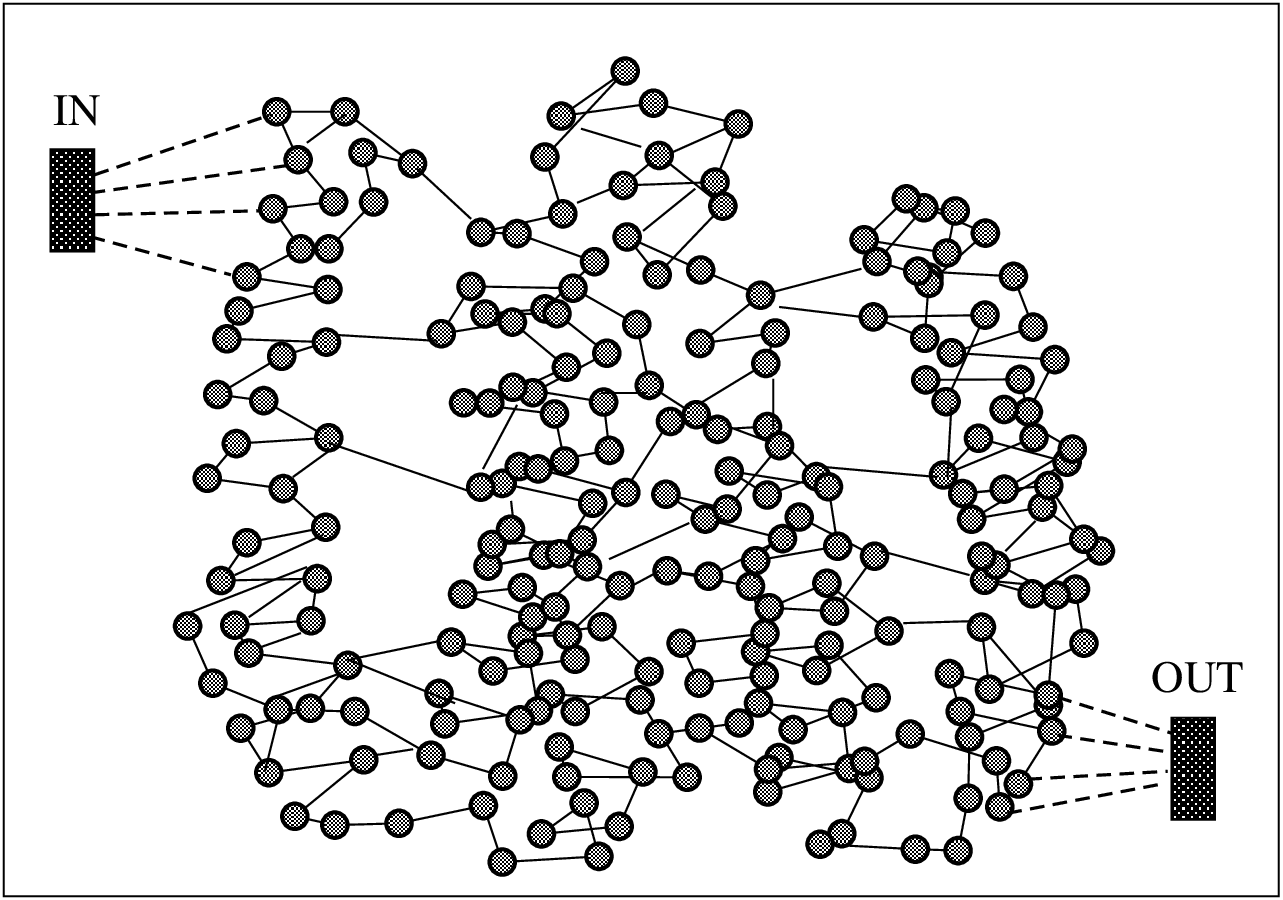}
                \caption{Drawing of a graph corresponding to a single protein. The interaction radius is taken small as a few {\AA}ngstrom.}
        \label{fig:3dnet}
\end{figure}
\begin{figure*}
\centering\noindent
\includegraphics[scale=0.6]{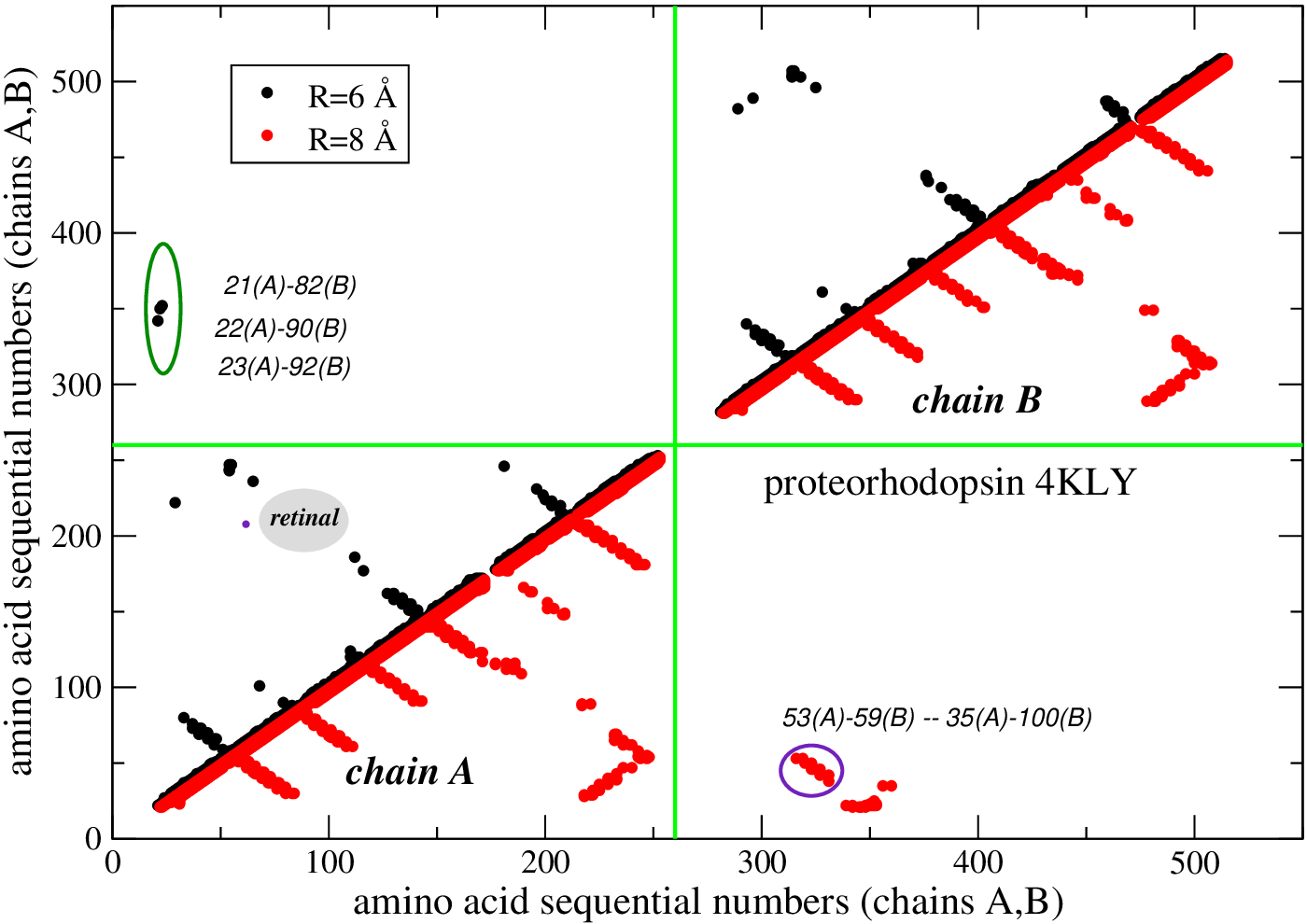}
\caption{Contact maps of the BPR mutant protein 4KLY, chains A\& B. Two different $R_{\rm c}$
values are considered, 6 and 8 \AA. The closeness of the two protomers is
revealed by points inside the I and IV quadrants. Inside the same protomer,
the closest helices appear to be A and B, D and E, F and G. The main contacts between the
 monomers are reported in the II and IV quadrants. }
\label{fig:figure2}
\end{figure*}
\par
The crystallographic investigation on BPR allows us to be acquainted on the differences
and main features of these proteins  and is also useful for analyzing
them within a computational method called impedance network protein analogue
(INPA) \cite{Alfinito08,Alfinito09,Alfinito09b,Alfinito11,Alfinito13}, able to correlate electrical and topological properties of proteins.
This method has been used to model the single protein properties; the results can be applied for modeling large samples  by a simple rescaling procedure \cite{Alfinito11}. 

\section{Modeling}
The analysis of the protein  topological and electrical properties is performed by considering only the backbone of the protein.  The interactions between the amino acids  of the backbone are mimicked by means of an impedance network.
In particular, each node of the network corresponds to a single carbon-$\alpha$ of the protein amino acids and the interactions between amino acids are taken as responsible of the charge transfer and/or the charge polarization.
To this aim, the links between couples of nearest neighboring nodes reproduce elementary impedances 
(resistance and capacitor in  parallel) able to capture the electrical response of the protein under different experimental conditions \cite{Alfinito13}. 
As matter of fact, many recent experiments \cite{Hou06,*Hou07,Jin06, Casuso07,Melikyan11}  evidenced electrical responses in sensory proteins used as active parts of two terminals electronic devices (micro-  and nano- electrical structures, functionalized substrates, protein anchoring on field effect transistors, etc). 
By construction, INPA models the electrical characteristics of a single protein as a result of electrical interactions in a specific protein configuration. A change in the protein configuration implies a change in the protein electrical response. 
Details on this model have been previously reported, anyway for the reader's
convenience, here we recall its mean features. 
In a first step, the protein is mapped into a graph like that drawn in  Fig.~ \ref{fig:3dnet}, where the existence or not of a link between a couple of nodes is dictated by their  distance: it has to be less than the assigned interaction radius, R$_{\rm c}$. Finally these links are associated with an elementary impedances, $Z_{i,j}$:
\begin{equation}
Z_{i,j}={l_{i,j}\over {\mathcal{A}}_{i,j}}   
\frac{1}{(\rho^{-1} + {\rm i} \epsilon_{i,j}\, \epsilon_{\rm 0}\omega)}  
\label{eq:1}
\end{equation}   
where ${\mathcal{A}}_{i,j}=\pi ({R_{\rm c}}^2 -l_{i,j}^2/4)$, is the cross-sectional
area between two spheres of radius R$_{\rm c}$ centered on the $i$-th and $j$-th node, respectively;
$l_{i,j}$ is the distance between these centers, $\rho$ is the 
resistivity, taken to be the same for every 
amino acid, and allowed to change with the intensity of applied bias \cite{Alfinito11}, 
${\rm i}=\sqrt{-1}$ is the imaginary unit,  
$\epsilon_0$ is the vacuum permittivity,
and  $\omega$ is the circular frequency of the applied voltage. 
The relative dielectric constant of the couple of $i,j$ amino 
acids, $\epsilon_{i,j}$, is expressed in 
terms of the intrinsic  polarizability of the $i,j$ amino acids.
By positioning the input and output electrical contacts on conveniently chosen sets of nodes, the network is solved within a  Kirchhoff scheme, for an applied bias value.
\par 
In DC condition, this network produces a structure dependent current voltage (I-V) characteristic. 
To account for the super-linear current at increasing applied voltages, a tunneling mechanism of charge transfer is included by using a stochastic approach within a Monte Carlo scheme \cite{Alfinito11}. 
In particular, following the Simmons model \cite{Simmons63}, confirmed by data on bR \cite{Jin06, Casuso07}, a mechanism containing two different tunneling processes, a direct tunneling (DT) at low bias, and a Fowler-Nordheim tunneling (FN) at high bias,  is introduced. 
The tunneling mechanism, in the present framework, aims to explain the change in resistance, which becomes bias-dependent. 
Therefore, the resistivity value of each link is chosen between a low
value $\rho_{\rm min}$, taken to fit the current at the highest voltages, and a high value $\rho(V)$, which depends on the voltage drop between network nodes as:
\begin{eqnarray}
\rho(V)&=& \rho_{\rm MAX}  \qquad \qquad (eV \le \Phi),  \label{eq:3a}\\
\rho(V)&=&\rho_{\rm MAX}(\frac{\Phi}{eV})+\rho_{\rm min}(1- \frac{\Phi}{eV}) \quad   (eV \ge  \Phi) 
\label{eq:3b}
\end{eqnarray}
where $\rho_{\rm MAX}$ is the maximal resistivity value taken to fit the I-V characteristic at the lowest voltages (Ohmic response)  and $\Phi$  is the height of the tunneling barrier between nodes.
The transmission probability of each tunneling process is given by:
%

\begin{eqnarray}
 P^{\rm DT}_{i,j} &=& \exp \left[- \alpha \sqrt{(\Phi-\frac{1}{2}
eV_{i,j})} \right] \quad  
 (eV_{i,j}  \le \Phi) , \label{eq:4a} \\
{P}^{\rm FN}_{ij}&=&\exp \left[-\alpha\ \frac{\Phi}{eV_{i,j}}\sqrt{\frac{\Phi}{2}} \right] 
 \qquad   (eV_{i,j} \ge \Phi)  
\label{eq:4b}
\end{eqnarray}
%
%
where $V_{i,j}$ is the potential drop between the couple of $i,j$ amino acids, $\alpha =\frac{2l_{i,j}\sqrt{2m}}{\hbar}$, and $m$  is the electron effective mass, here taken the same of the bare value.
\section{Results}
In the following the INPA model is used to carry out a comparative investigation among different photo-receptors, in order to evidence their similarities and  main differences.
\subsection{Topological properties}
The topological features of a single BPR (protomer) have been analyzed by using its contact map. 
This consists of a 2D representation of the interactions among amino acids: each link is described by a couple of numbers corresponding to the sequential numbers of the connected amino acids. 
Therefore, in the plane of these coordinates, each link corresponds to a single point.
In this representation, the picture of the links: (i) signals the degree of connection
of the protein, (ii) shows the regions of maximal connectivity, (iii) reveals the topological transformations \cite{Alfinito09b}. 
The contact maps are symmetric under the reflection around the diagonal,
therefore, two different set of data, in the same graph, one on the left side of the diagonal, the other on the right side of the diagonal \cite{Alfinito09b}  have been reported. 
Both these datasets correspond to the same monomer. 
In this paper we test, for the first time, the use of contact
maps for detecting the protein \textit{quaternary structure}, i.e. the interactions among different monomers in the same protein. 
In particular, Fig.~\ref{fig:figure2} reports the contact maps of two different monomers, say the chains A and B of the BPR 4KLY. 
The interactions are described for two different values of $R_{\rm c}$, say 6 and 8 \AA. 
In doing so, we produce a contact map in 4 quadrants, enumerated in clockwise direction from the bottom left. 
The diagonal quadrants report the contact maps of each monomer with itself, for two different $R_{\rm c}$ values. 
The quadrant II reports the interactions between the two chains for $R_{\rm c}=8$ \AA, and finally, the quadrant IV reports the interaction between the two chains for $R_{\rm c}=6$ \AA.
\par
The contact maps for the two chains are quite similar. For both chains, the most connected region is the one containing the retinal. Furthermore, it is also possible to observe some specific links between the two structures, emphasized by ellipses, which are typical of the quaternary structure of these new proteins \cite{Ran13}. This result seems to confirm  the biochemical interactions are mainly short-ranged and different parts of different monomers interact only if they are close.
 %
\subsection{AC response}
\begin{figure}
        \centering
        \includegraphics*[width=0.4\textwidth]{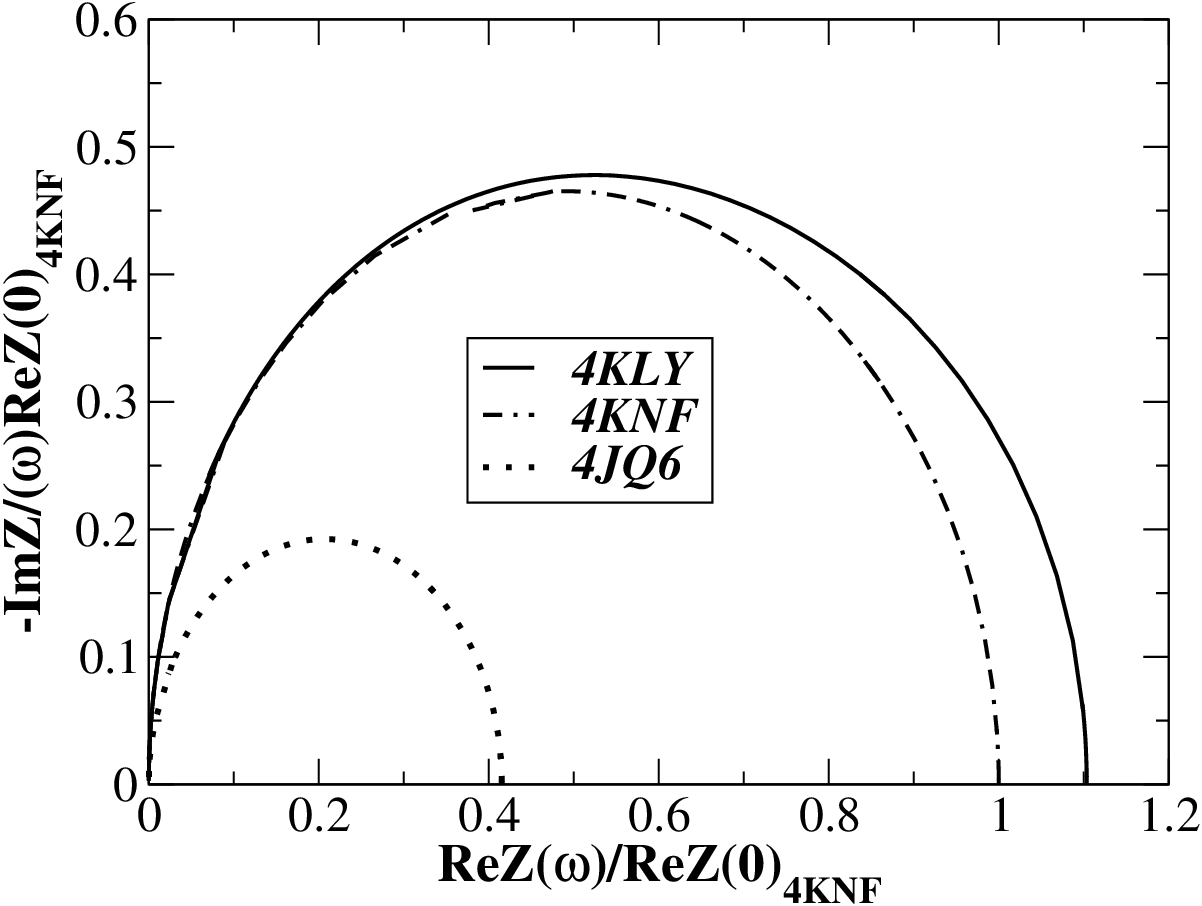}
                \caption{Nyquist plots for the three blue light proteorhodopsins. The interaction radius is $R_{\rm c}$=~8 \AA. Both real and imaginary parts of all the plots have been normalized to the value of the zero-frequency impedance of 4KNF.}
        \label{fig:fig3}
\end{figure}

\begin{figure}
        \centering
                \includegraphics*[width=0.40\textwidth]{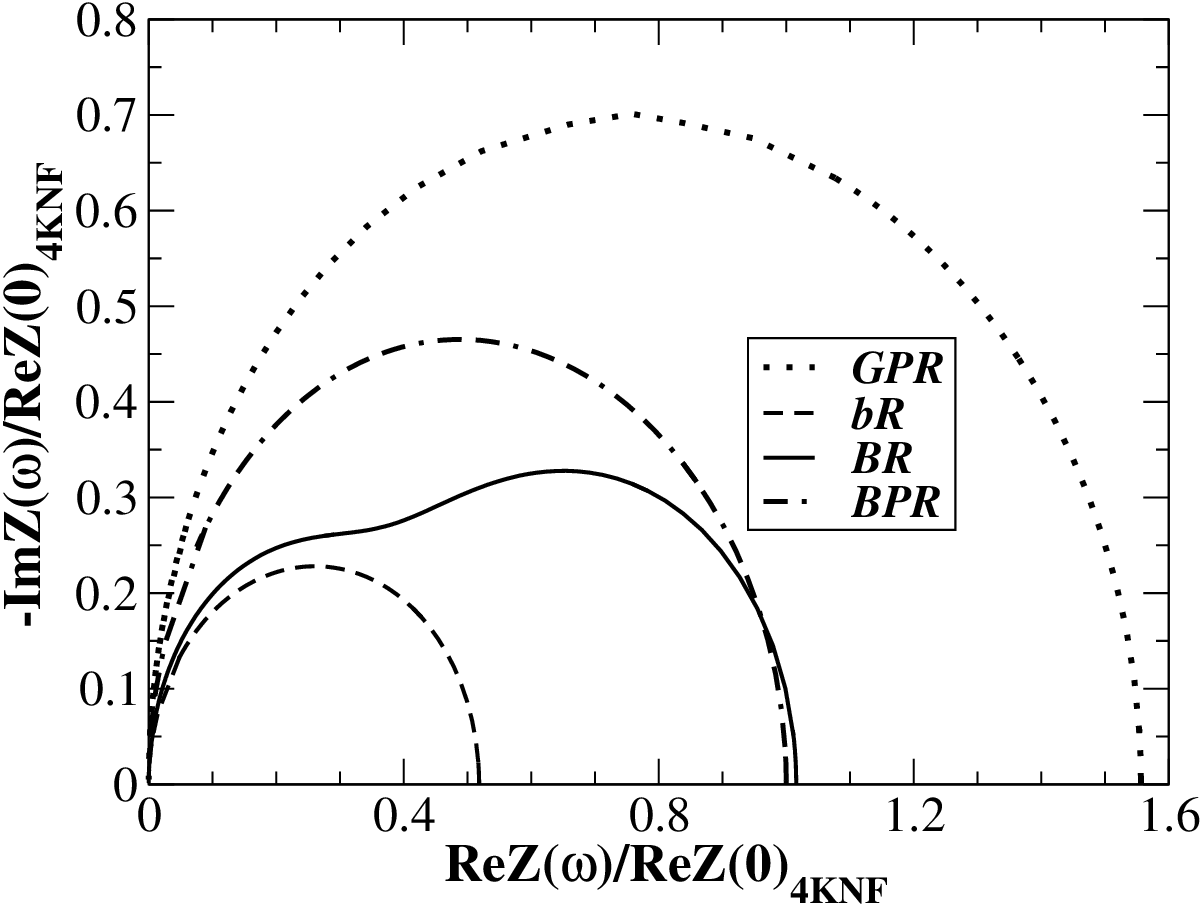}
        \caption{Nyquist plot of four different opsins, $R_{\rm c}$=~8 \AA. Both real and imaginary parts of all the plots have been normalized to the value of the zero-frequency impedance of 4KNF.}
        \label{fig:fig4}
\end{figure}

\begin{figure}
        \centering
                \includegraphics*[width=0.40\textwidth]{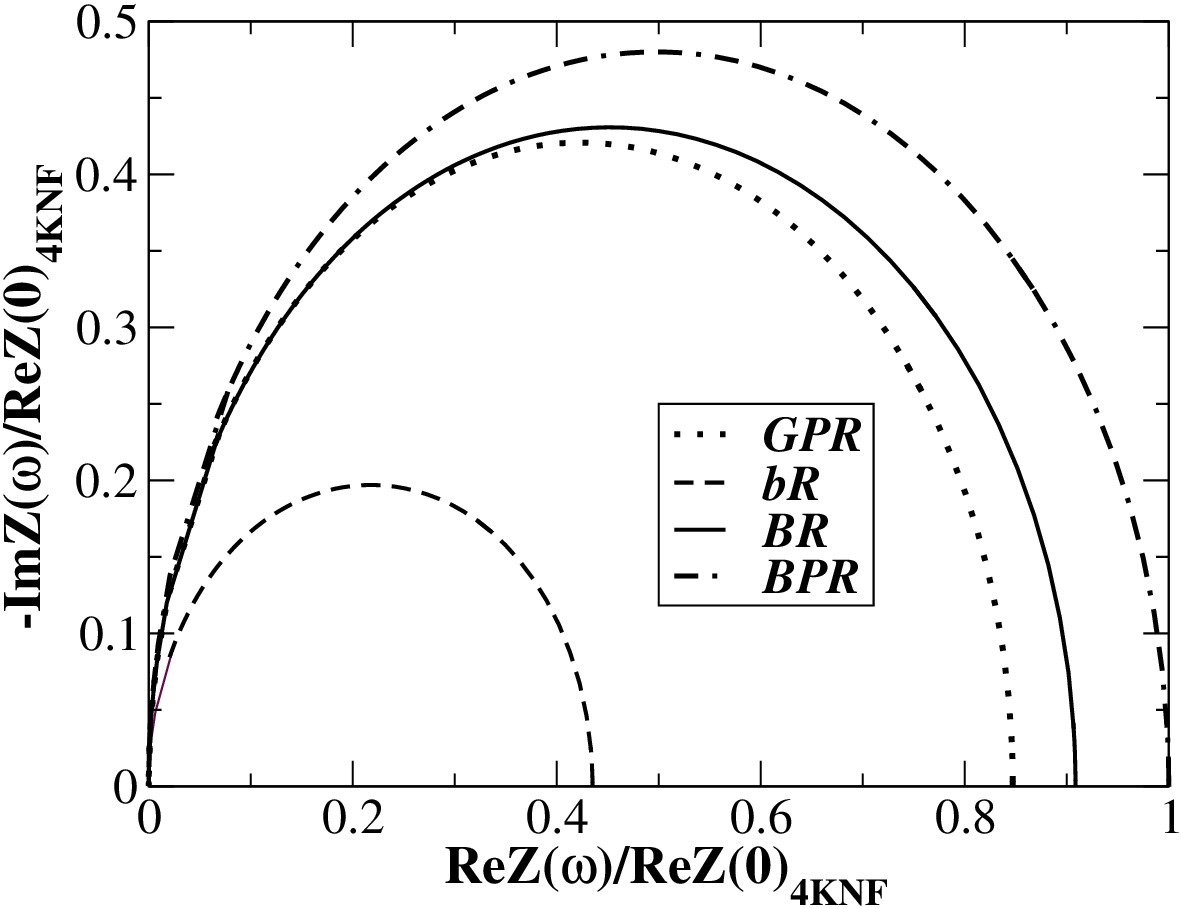}
        \caption{Nyquist plot of four different opsins, $R_{\rm c}$=15 \AA. Both real and imaginary parts of all the plots have been normalized to the value of the zero-frequency impedance of 4KNF.}
        \label{fig:fig4a}
\end{figure}

 The Nyquist plots, often used to report the electrochemical impedance spectroscopy (EIS) experiments \cite{Hou06, *Hou07}, can be also calculated inside the INPA approach, by using Eq.~\ref{eq:1}. 
In particular, these graphs plot the protein global imaginary impedance vs the real part, at different values of the applied bias circular frequency, $\omega$. 
In general, see  Fig.~\ref{fig:fig3}, the shape of the BPR plots resemble a perfect semicircle, as  given by a single R-C parallel circuit, i.e. the different elementary impedances of the network work in synchrony, which, on the other hand, means that the interactions percolate the whole protein and there are no bottleneck or dead-ends. 
We can also observe the  role of the protein length: the smallest protein, 4JQ6 has the smallest global resistance, $Z(0)$. 
On the other hand, the protein length is not the only factor determining the shape of the Nyquist plot: all the graphs have been normalized to the $Z(0)$ value of the longest protein, 4KNF, and it is clear  the largest resistance value is that of 4KLY, which is a little bit smaller than 4KNF. 
In conclusion, not only the protein length but also its 3D arrangement determines the AC electrical responses.
As a further investigation, the BPR-4KNF has been compared with bR, BR and GPR,  
Fig.~\ref{fig:fig4}. 
In this case it is found that the protein length has only a partial role in the AC response: for instance, BR and BPR which differ for about 100 amino acids have a very close value of $Z(0)$. 
Furthermore, the semicircle that pertains to the simple R-C circuit is not recovered by BR, which, instead, shows two markedly different response times. 
This kind of behaviour signals the lack of homogeneity of the network, due to the very small value of the interaction radius \cite{Alfinito08, Alfinito09b}. 
By increasing  the interaction radius we force a single global response of the protein, recovering a perfect semicircle (see Fig.~\ref{fig:fig4a}). 
A simple interpretation of the  protein electrical response is linked to the specific protein topology  : inside the protein the amino acid distribution is not uniform, instead helices (or planes where present) constitute homogeneous domains which do not coordinate at small values of the interaction radius: increasing $R_{\rm c}$, they can synchronize, giving a collective response.   
Enlarging the interaction radius, the number of connections among different domains grows and a single global response appears. 
On the other hand, this is equivalent to affirm that the secondary structure prevails over the tertiary structure and that possible conformational changes modify only the latter. 
In some sense, this parallels the dynamics of magnetic systems which, under the action of an external magnetic field (here the electric field), deforms the domain walls. The detectable effect is   a finite magnetization. 
In the present case, the conformational change has a measurable electrical counterpart in a photocurrent \cite{Alfinito13b,*Alfinito13c} which is tuned on/off by switching the electromagnetic field. 
\par
As a final comment, looking at Fig.~\ref{fig:fig4a}, we notice  the non monotonic impedance variation as function of $R_{\rm c}$. 
This signals the non-homogeneity in the amino acid distribution and has been also observed in different protein receptors \cite{Alfinito09b}. 
Concerning this point, the widest variation comes from GPR, whose resistance is 60\% larger than that of 4KNF,  at $R_{\rm c}$=8~\AA \ and 15\% smaller than that of 4KNF at $R_{\rm c}$=15~\AA.  
This means that the different domains inside the protein have reciprocal distances large at least as 8~\AA, thus by increasing $R_{\rm c}$, the resistance abruptly slows down. 
\subsection{Current-Voltage characteristics}
\begin{figure}
        \centering
                \includegraphics*[width=0.40\textwidth]{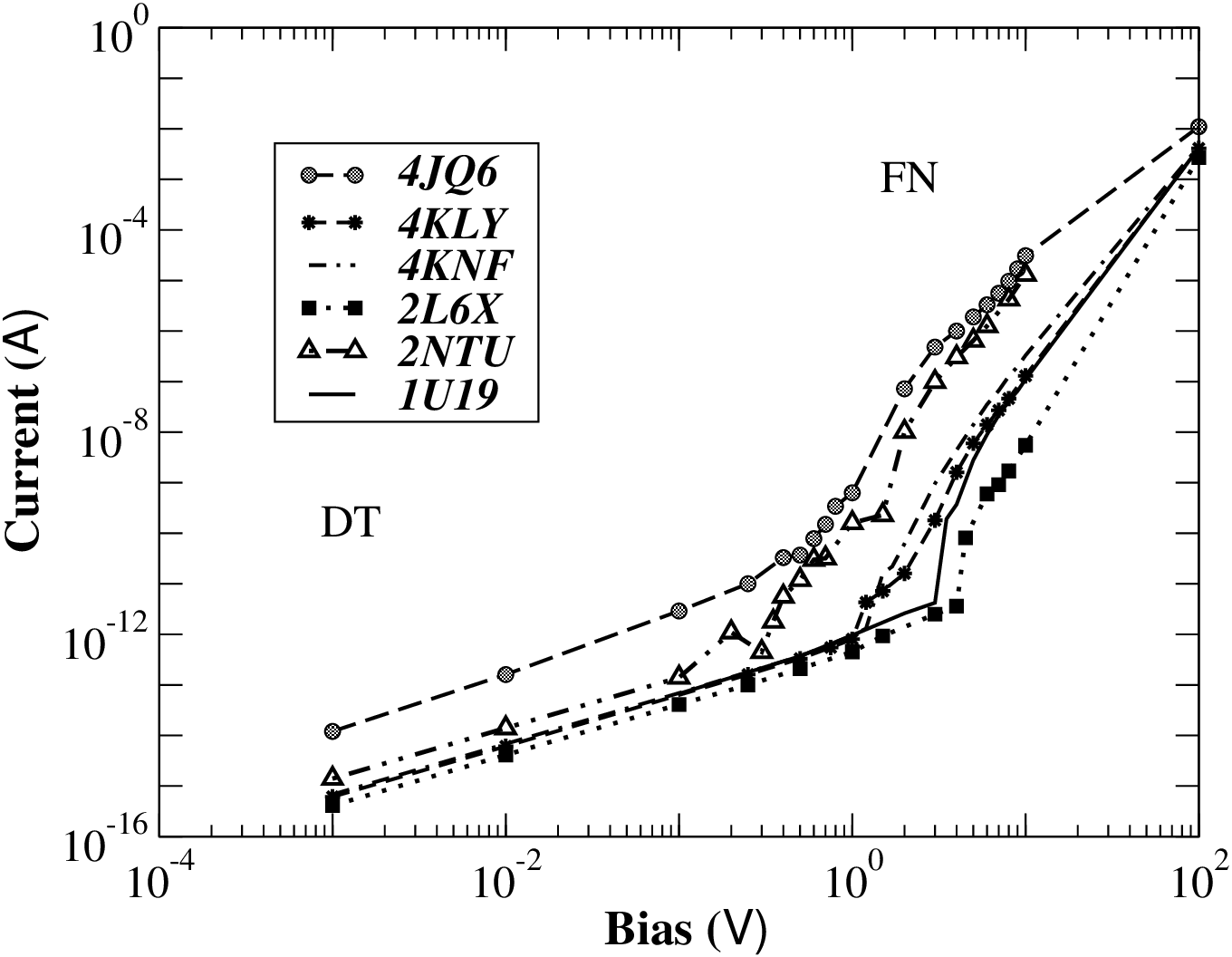}
        \caption{I-V characteristics for the single protein, $R_{\rm c}$=8 \AA.}
        \label{fig:currentnew}
\end{figure}
\begin{figure}
        \centering
                \includegraphics*[width=0.40\textwidth]{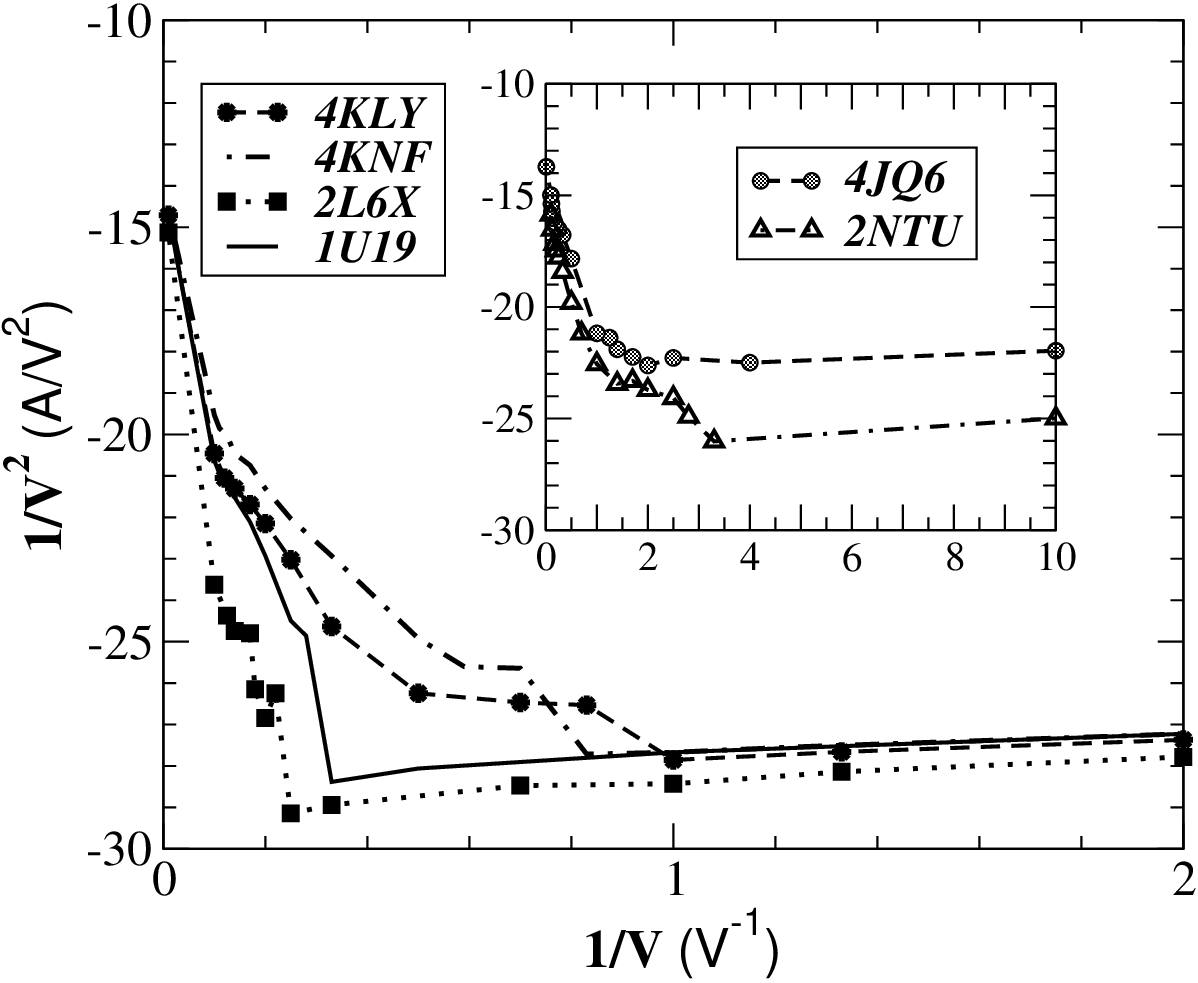}
        \caption{Fowler representation of a set of current-voltage values.
        The values reported in the axes are chosen to emphasize the FN regime at high voltages}
        \label{fig:fowler}
\end{figure}
\begin{figure}
        \centering
                \includegraphics*[width=0.40\textwidth]{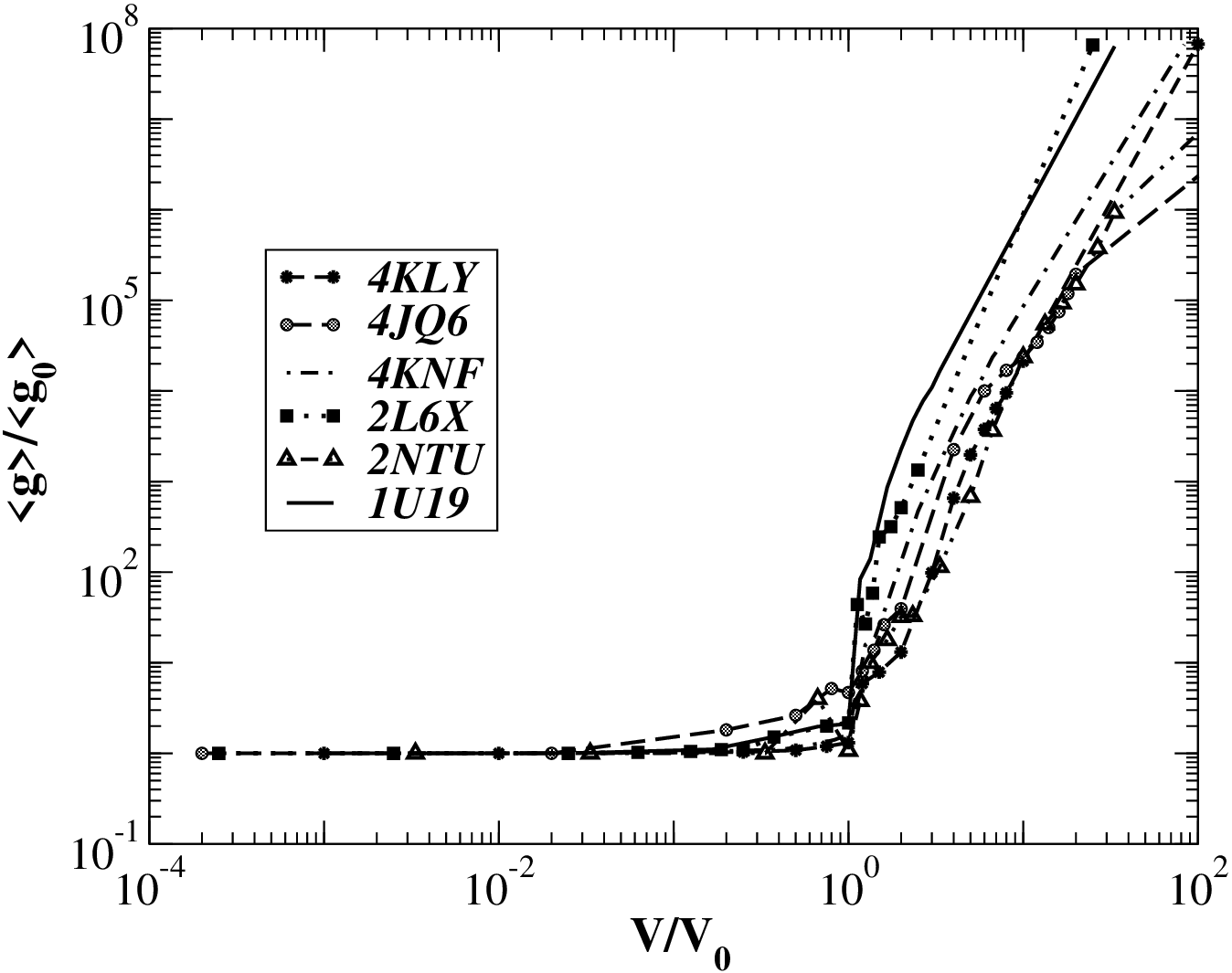}
        \label{fig:conductance}
\caption{Single protein conductances mean values $\langle g\rangle$. Values are normalized to the Ohmic asymptote $\langle g_{\rm 0}\rangle$. $V_{\rm 0}$ indicates the transition bias value for each protein.}
\end{figure}
Both GPR and bR show a relevant current response under the
application of a zero-frequency voltage. 
In particular, they exhibit a region of quasi-linear behavior with a resistivity similar to that of a medio-gap semiconductor, followed by a super-linear current at increasing voltages. 
The I-V characteristic appears quite symmetric under
the polar inversion of the bias value. 
Furthermore, both proteins exhibit an increase of the current intensity in the presence of a visible light \cite{Jin06,Melikyan11}.
\par
So far, the interpretation of experimental results, made in the framework
of the INPA, model has been satisfactory when applied to bacteriorhodopsin
(bR). 
As a matter of fact, most of I-V measurements have been conducted
on monolayers of this protein
Therefore, it has been possible to calibrate the network parameters in order to reproduce on a very wide range (more than 6 orders of magnitude) the current responses. 
In particular the values: R$_{\rm c}$= 6 \AA \,, $\Phi$= 0.219 meV, $\rho_{min}=4\times 10^{5}\Omega$ \AA\ and $\rho_{min}=4\times 10^{13}\Omega$ \AA\  have been found to give the best agreement.
\par
The main objective of the present investigation is to compare the behaviour of different proteins within the same conditions, therefore all the calculations have been performed by using the previous parameters. Only the value of the interaction radius has been changed in
  $R_{\rm c}$=8 \AA \ .
As a matter of fact, present PDB entries of BPR are quite incomplete, i.e. there are many missed amino acids, and this  produces large vacancies in the network and finally not connected networks for too small values of  $R_{\rm c}$.
At first glance, all the calculated  I-V characteristics reported in Fig.~\ref{fig:currentnew} show a double regime, the quasi-linear or direct (DT) regime and the super linear or Fowler-Nordheim regime (FN). 
The crossing region is between 0.3V and 4.0V for all the opsins.
The differences between the single protein current values are quite large, in the DT they are of about 3 orders of magnitude and this value seems not directly related to the protein length, although it appears the two smallest proteins have the higher conductivity (see Tab.~\ref{tab:Conductance}).
\begin{table}[h]
        \centering
                \begin{tabular}{lcclccl}
                \hline\hline
                        {\it Protein} & && {\it $\langle g_{\rm 0} \rangle$(pS)}& & & Amino acid \\ &&& &&& number\\
                        \hline
                        1U19& &&0.67 & & & 348 \\
                        2NTU & && 1.4 &&& 222\\
                        2L6X & && 0.41 &&& 235 \\
                        4QJ6 & && 16.0 &&& 203 \\
                        4KNF & && 0.68  &&& 231\\
                        4KLY & && 0.61  &&& 228\\
                        \hline\hline
                \end{tabular}
        \caption{Conductance of single protein as calculated by INPA model with $R_{\rm c}$=8 \AA. The number of amino acids is that given by the PDB sequences.}
        \label{tab:Conductance}
\end{table}

On the other hand, in the FN region, see Fig.~\ref{fig:currentnew}, the differences among the current values reproduce the differences in impedances observed in Fig.~\ref{fig:fig4}.
The origin of this particular current response is suggested to be the tunneling of electrons among localized states \cite{Zvyargin07}. 
In low bias conditions, convenient is to describe the tunneling barrier as rectangular;  the increasing bias is instead described as a  barrier shape deformation, from rectangular to triangular \cite{Simmons63}. 
The former condition, direct tunneling (DT), is associated to a low, quasi-linear current growth, the latter , injection or Fowler-Nordheim regime(FN), describes a high, super-linear responses.

To reach the FN regime very high potential values are needed and thus very high current values  are attained.
Accordingly, it requires extreme skill in the experimental set-up \cite{Casuso07} and often only the DT regime is observed  \cite{Jin06,Melikyan11}. 
The way followed to get the FN regime has been to perform nanometric measurements i.e. onto nanometric sized samples and resolving  very low (from  0.3 picoampere to about 10 nanoampere) current intensities. 
Due to the Joule heating, larger samples do not allow this kind of measurements. 
Here we calculate the single protein I-V characteristics and this allow us to classify the opsins under considerations into proteins with a high transition bias value $V_{\rm 0}$, HTBV,  and proteins with a low transition bias value, LTBV,  proteins. 
Notice that the specific value of the transition bias is $R_{\rm c}$-dependent. 
In particular, taking $R_{c}=8$ \AA,  1U19, 2L6X and 4KNF, 4KLY, are HTBV while 4JQ6 and 2NTU are LTBV as shown in Fig.~\ref{fig:fowler}.   
Finally, these differences could drive the kind of experiments and the possible applications of the opsins under considerations: in order to avoing Joule superheating and also getting detectable current response,  the LTBV proteins should be used in very small samples, while the HTBV proteins should be used in large samples.
For both of them a light induced current variation is expected, although the bovine rhodopsin could show a decreasing of current intensity.
Concerning the origin of this photocurrent, the guess is that it is mainly due to the conformational change associated with photon absoprion. 
As a matter of fact, the retinal rearrangement which follows the photon absorption produces a modification of the 3D structure of the whole protein.
Finally, due to the different rearrangement of the retinal in these proteins \cite{Alfinito09}, a different kind of response for bovine rhodopsin
with respect to  bacteriorhodopsin is expected.
\par
As a final remark, we notice that an increased value of $R_{\rm c}$ changes the network degree of connections and, in turn, the value of the global current. 
In the case of bR 2NTU, an increase of $R_c$ from 6 to 8 \AA \ increases the lowest current of one order of magnitude and decreases the bias transition value from 2.5 to 0.3 V.
On the other hand, the main features of the current response remain the same.
\section{Conclusions}
In the framework of materials useful for electronics the coming in of new players, like opsins or more generally proteins, opens a new branch of molecular electronics, which, joining  proteomics (i.e. the large-scale study of proteins) with electronics is coined as \textit{proteotronics}.
The technological impact of proteotronics is impressive, going from the diagnostic and medical applications \cite{Lee12, Lima05,*Deisseroth10, Hegemann10, Humayun96, Ran13, *Saeedi11, Ghezzi11,*Ghezzi13, Ahuja11} to the development of new energy power supplies \cite{King12,Renu14}. 
For these reasons, many and different experiments are underway, with the aim of producing large-scale usable devices.
Furthermore, the theoretical interpretation of experiments can take advantage of a computational model, called INPA, used to describe and predict structural and electrical properties of proteins.
Accordingly, in this work the INPA model has been  applied to a set of different opsins and it revealed the difference in electrical response  through the differences in structures. 
Furthermore, the specific quaternary structures of the BPRs has been correlated to  the short-range interactions between monomers. 
The electrical investigation has covered the linear response (Ohmic region) and its frequency dependence, evidencing analogies and substantial differences among the proteins, partially due to the protein length.
The static I-V characteristics, with bias in the region $10^{-3} - 100 V$,  are found to be quite similar for the set of proteins considered here, evidencing the crossover of two different tunneling regimes of charge transfer between amino acids, direct tunneling at low voltages and indirect tunneling at high voltages.
On the other hand, it has been possible to identify two different classes of proteins, those showing a slow transition with a very low critical bias value and those showing  a sharp transition, with high critical bias value. This suggests different applicative uses for the opsins.

The INPA model is today in progress and will take advantage of the discovery of new effects in experiments, like those resulting from ref.\cite{Renu14}.  

As a final remark, we emphasize the most amazing perspective of the physical investigation on proteins: they have a similar but not identical behaviour and the differences arise from both structural and functional properties. In particular, quite similar proteins, belonging to the same class, like 4JQ6 and 4KLY  show different electrical responses and quite different proteins like 1U19 and 2L6X show similar behaviour. This means that much on this fields has to be made, to get a deep understanding of the phenomena  and avoid misuses of this full-of potentiality material.
\begin{acknowledgments}
{This research is supported by the European Commission under the Bioelectronic Olfactory Neuron Device (BOND) project within the grant agreement number 228685-2.}
\end{acknowledgments}

\end{document}